\documentclass[12pt,A4]{article}
\usepackage{amssymb, amsmath, enumerate}
\begin{document}
\newtheorem{thm}{Theorem}
\newtheorem{lm}{Lemma}
\newtheorem{prop}{Proposition}
\newtheorem{cor}{Corollary}
\newtheorem{defi}{Definition}
\newtheorem{eg}{Example}
\newtheorem{coj}{Conjecture}
\newtheorem{rem}{Remark}

\numberwithin{equation}{section}

\newcommand{\B}{{\bf Proof\ } }
\newcommand{\A}{$\blacksquare$ }
\newcommand{\CC}{\mathcal C}
\newcommand{\CF}{\mathcal F}
\newcommand{\CR}{\mathcal R}
\newcommand{\CV}{\mathcal V}
\newcommand{\ch}{\mathcal H}
\newcommand{\ca}{\mathcal A}
\newcommand{\cb}{\mathcal B}
\newcommand{\cp}{\mathcal P}
\newcommand{\cs}{\mathcal S}
\newcommand{\X}{\mathcal X}
\newcommand{\W}{\mathcal W}
\newcommand{\ic}{\mathcal I}

\newcommand{\M}{\mathcal M}
\newcommand{\mco}{\mathcal O}
\newcommand{\cl}{\mathcal L}
\newcommand{\C}{\Bbb C}
\newcommand{\N}{\Bbb N}
\newcommand{\R}{\mathbb{R}}
\newcommand{\Q}{\mathbb{Q}}
\newcommand{\I}{\mathbb{I}}
\newcommand{\T}{\mathbb{T}}
\newcommand{\CP}{\mathbb{CP}}
\newcommand{\h}{\mathfrak{H}}
\newcommand{\dia}{\mathrm{dia}}
\newcommand{\id}{\mathrm{id}}
\newcommand{\1}{\mathbf{1}}
\newcommand{\mo}{{}^{\scriptscriptstyle -1}}

\newcommand{\rf}{{}_{\scriptscriptstyle(F)}}
\newcommand{\lf}{_{}{\scriptscriptstyle F}}
\newcommand{\eps}{\epsilon}
\newcommand{\Del}{\Delta}
\newcommand{\del}{\delta}
\newcommand{\Lan}{\Lambda}
\newcommand{\Gam}{\Gamma}
\newcommand{\gam}{\gamma}
\newcommand{\ten}{\bigotimes}
\newcommand{\op}{\oplus}
\newcommand{\im}{\imath}
\newcommand{\st}{\otimes}
\newcommand{\fl}{\rightarrow}
\newcommand{\mpt}{\mapsto}
\newcommand{\ti}{\times}

\newcommand{\Om}{\Omega}
\newcommand{\sfi}{\varphi}
\newcommand{\om}{\omega}
\newcommand{\lan}{\lambda}
\newcommand{\sig}{\sigma}

\newcommand{\al}{\alpha}
\newcommand{\be}{\beta}
\newcommand{\te}{\theta}
\newcommand{\all}{\forall}
\newcommand{\ci}{\circ}
\newcommand{\sm}{\shortmid}
\newcommand{\ld}{\ldots}
\newcommand{\cd}{\cdots}
\newcommand{\p}{\partial}
\newcommand{\n}{\nabla}
\newcommand{\mt}{\mapsto}
\newcommand{\se}{\subseteq}
\newcommand{\fa}{\forall}
\newcommand{\iy}{\infty}
\newcommand{\tl}{\textless}
\newcommand{\tg}{\textgreater} 
\newcommand{\tel}{\textless} 

\newcommand{\teg}{\textgreater} \newcommand{\eqn}[2]{\begin{equation}#2\label{#1}\end{equation}}

\textheight 23.6cm \textwidth 16cm \topmargin -.2in \headheight
0in
\headsep 0in \oddsidemargin 0in \evensidemargin 0in \topskip
28pt
\title{
\textbf{Reformulation of Quantum Theory}}

\author{Seyed Ebrahim Akrami\footnote{Email: akramisa@ipm.ir,
Skype: Seyed Ebrahim Akrami}
\\{\small 
(Semnan University, Semnan, Iran)}}
\date{January 1, 2022}
\maketitle
\begin{abstract}
We first recall a fact which is well-known  among mathematical physicists although lesser-known among theoretical physicists that the standard quantum mechanics over a complex Hilbert space, is a Hamiltonian  mechanics, regarding the Hilbert space as a linear real  manifold equipped with its canonical symplectic form and restricting only to the expectation-value functions of Hermitian operators. Then in this framework, we reformulate the structure of quantum mechanics 	in the language of symplectic manifolds and avoid linear structure of Hilbert space in such a way that the results can be stated for an arbitrary symplectic manifold. 
\end{abstract}
\section{Introduction}
``God does not play dice with the Universe", ``I want to know God's thoughts".  Einstein.

Newton postulated in the  Principia, the mathematical principles of Nature and his followers Lagrange and Hamilton reformulated the Principia in the form of Lagrangian and Hamiltonian mechanics while keeping its spirit. Then Maxwell and Einstein reformulated and completed the Principia while keeping its spirit again, made serious modifications in the principles like abandoning absolute space and time and  far-distance-acting forces. Then the new generation of physicists made serious changes in the spirit of Principia and new Principia named Quantum Mechanics was postulated by  Dirac, von Neumann and others for mathematical principles of Nature. But as Einstein predicted, the new Principia, although very excellent in describing Nature, was not complete. Also, the new Principia is not reconciled with the old Principia; we refer to the problems of classical limit and quantum gravity.

There are very radical changes in the spirit of old Principia within the new Principia. In the new Principia unlike the old one, the space of states of a system of finitely-many particles are infinite-dimensional. This article is an attempt in reformulating  Quantum Mechanics in the spirit of old Principia, i.e. Classical Mechanics.
In section 2, we bring some preliminaries from symplectic manifolds and Hamiltonian mechanics and recall that quantum mechanics over a complex Hilbert space is a Hamiltonian mechanics regarding the  Hilbert space as a real manifold under its canonical symplectic form and restricting ourselves to expectation-value functions. Also, in this section by the aid of the later mentioned fact, we reformulate quantum mechanics in such a way that we avoid its linear, complex and infinite-dimensional structures. Next, in section 3, we generalize the result obtained in previous section to a general Hamiltonian mechanics and thus introduce new principles of the Principia. We will introduce a new concept named \textbf{quantum function} over phase space which is going to be a replacement for the Hermitian operators of standard quantum mechanics. In section 4, we recover the standard quantum mechanics from our theory and 
finally at section 5, we will impose a fundamental equation to produce quantum functions from given  quantum operators. The essential task then will be to solve this equation and translate the well-known Born rule of standard quantum theory to our theory.
\section{Mathematical Statement of the Problem and the Strategy to its Resolution}
Here, we recall some concepts and facts on symplectic manifolds and Hamiltonian mechanics. For details, see pages 42-46 of  \cite{T}, section 9.1 of \cite{AMR} and chapters 2,5 and 10 of \cite{MR}.

A symplectic form $\om$ over a real manifold $M$ is a (strongly) non-degenerate closed $2$-form over $M$. We can assign to any function $f$ a vector field $X_f$ called \textbf{Hamiltonian vector field} which is the unique vector field satisfying
\eqn{}{\om(X_f,Y)=df(Y)} for any vector field $Y$. The \textbf{Hamilton equation} for a given function $f$ is
\eqn{dotxi}{\dot{\xi}=X_f(\xi(t))}$\xi(t)\in M$.
We also have a Poisson bracket
\eqn{fg}{\{f,g\}=X_fg=\om(X_f,X_g)} for functions $f$ and $g$, where $X_fg=dg(X_f)$ is the derivative of $g$ at direction $X_f$.  
A  \textbf{Hamiltonian dynamical system} or \textbf{Hamiltonian mechanics}  is a triple $(M,\om,H)$ including a real (finite or infinite dimensional) manifold $M$, a symplectic form $\om$ (a non-degenerate,  closed $2$-form over $M$) and a function $H$ over $M$, called  Hamiltonian.

A very important and essential fact which is lesser-known among theoretical physicists and however is well-known among mathematical physicists like Abraham, Marsden and Ratio, is that a quantum mechanics over a complex Hilbert space is in fact an example of a Hamiltonian mechanics  when we regard the Hilbert space as a real manifold equipped with its canonical symplectic form which provides a Poisson bracket for the real functions (not operators) over the Hilbert space. The Poisson bracket of expectation-value functions of two Hermitian operators  coincides with the expectation-value function of the commutator of the operators. This viewpoint is different than the view of theoretical physicist Dirac and mathematical physicists Von Neumann and  Takhtajan, \cite{T}, in which quantum mechanics is regarded as a Poisson algebra of Hermitian operators not as a symplectic or Poisson manifold. 

We first recall that the canonical symplectic form of $\R^n\ti\R^n$ can be translated to the symplectic form $\Om(v,w):=-2\mathrm{Im}\textless v|w\textgreater$ over $\C^n$.

\begin{thm}\label{1} 
	Quantum mechanics over   a complex (finite or infinite dimension) Hilbert space $\ch$ is an example of a  Hamiltonian mechanics    regarding $\ch$ as a real manifold, under the following symplectic form
	\eqn{}{\Om(|v\teg,|w\teg):=-2\hbar\mathrm{Im}\textless v|w\textgreater,~~~~~~~~\fa |v\teg,|w\teg\in \ch.} To each Hermitian linear operator $A$ over $\ch$ we associate a real function $\tel A\teg:\ch\fl\R$
	\eqn{}{|\psi\teg\mt\tel\psi| A|\psi\teg.} The associated Hamiltonian vector field satisfies in
	\eqn{}{X_{\tel A\teg}=-\frac{i}{\hbar}A.}
	 The Poisson bracket of two functions $\textless A\textgreater$ and $\textless B\textgreater$, where $A$ and $B$ are two Hermitian operators, satisfies in 
	\eqn{}{\{\textless A\textgreater,\textless B\textgreater\}=\textless [A,B]\textgreater.}
	The Hamilton equation  for the Hamiltonian $\tel\mathbb{H}\teg$ is nothing other than the Schrodinger equation
	\eqn{}{i\hbar\dot{\psi}=\mathbb{H}\psi}for a given Hamiltonian operator $\mathbb{H}$ .
	\end{thm}
For the proof see, \cite{AMR,MR}. We call $\ch$ with the above mentioned canonical symplectic from, as \textbf{canonical Hilbert-symplectic manifold}. By this theorem in fact, we reformulated quantum mechanics in the language of classical mechanics.  In below, we continue this program.
\begin{thm}
	Let $\ch$ be a complex Hilbert space  and $\{|\psi_n\teg\}$ be an orthonormal basis for $\ch$. We construct \textbf{coordinate functions}
	\eqn{}{u_n|\psi\teg:=\tel\psi_n|\psi\teg.} Then for any Hermitian linear operator $A$, whose eigenvectors are $|\psi_n\teg$, i.e. $A|\psi_n\teg=a_n|\psi_n\teg,$ we have
	\eqn{A}{\boxed{\textless A\textgreater=\sum_n a_n|u_n|^2},}
	\eqn{ihbarAun}{\boxed{i\hbar\{\textless A\textgreater,u_n\}=a_nu_n}} and
	\eqn{unxim1}{u_n(\xi_m)=\del_{mn},~~~~~~~~~~\textless A\textgreater(\xi_n)=a_n.}
\end{thm}
\B (\ref{A}) and (\ref{unxim1}) are obvious and well-known. But (\ref{ihbarAun}) is new which we now prove it. In Theorem \ref{1}, we recalled that  it is well-known that the Hamiltonian vector field  associated to the function $\textless A\textgreater$ is 
\eqn{}{X_{\textless A\textgreater}=-\frac{i}{\hbar}A.} Thus by (\ref{fg}), the linearity of $u_n$ and that $A$ is Hermitian, we have
\begin{eqnarray}
	i\hbar(\{\textless A\textgreater,u_n\})\psi&=&i\hbar d_\psi u_n(X_{\textless A\textgreater}(\psi))\nonumber\\&=&d_\psi u_n(A\psi))\nonumber\\&=&\textless \psi_n|A\psi\textgreater\nonumber\\&=&\textless A\psi_n|\psi\textgreater\nonumber\\&=&a_n\textless \psi_n|\psi\textgreater\nonumber\\&=&a_nu_n(\psi).\nonumber
\end{eqnarray} 
We give another proof. 
We recalled at Theorem \ref{1} that the Hamilton equation for the Hamiltonian $\textless A\textgreater$ is just the Schrodinger equation and thus its solutions are
\eqn{psit}{|\psi(t)\teg=\sum_nc_ne^{-\frac{iat}{\hbar}}|\psi_n\teg} for some initial state $|\psi_0\teg=\sum_nc_n|\psi_n\teg$. Now, it is enough to apply $u_n$ to the both sides of  (\ref{psit}) and use the following lemma. \A 
\begin{lm}For a given Poisson manifold $M$, a real function $f$ over $M$, a real number $a$ and a complex function $u$ over $M$, we have
\eqn{}{i\hbar\{f,u\}=au\Longleftrightarrow u(\xi(t))=e^{-\frac{iat}{\hbar}}u(\xi(0))} for all solutions   $\xi(t)$ of the Hamilton equation for the Hamiltonian $f$.	
\end{lm}
\B It is enough to note that $\frac{d}{dt}u(\xi(t))=d_\xi u(\dot{\xi})=\{f,u\}(\xi)$.
\A 

This theorem ruled out the linear, inner-product,  infinite-dimensional complex structures of quantum mechanics. Namely, it translated the structure of quantum mechanics to the language of symplectic real manifolds and Hamiltonian mechanics. Thus we can establish new principles for mechanics in the  language of classical mechanics which are translation of the principles of quantum mechanics. 

\section{New Principles for Quantum Mechanics}
\textbf{Principle 1.} The space of states  is a symplectic (or more generally a Poisson) real finite-dimensional manifold $M$.\\
\textbf{Principle 2.} An observable quantity is a real differentiable function $f:M\fl\R$  such that  there exist real numbers $a_n$, complex-valued functions $u_n:M\fl\C$ and states $\xi_n\in M$ satisfying
\eqn{ff}{\boxed{f=\sum_n a_n|u_n|^2},}
\eqn{ihbarfun}{\boxed{i\hbar\{f,u_n\}=a_nu_n},}
\eqn{sumnun2}{\sum_n|u_n|^2=1}
and
\eqn{unxim}{u_n(\xi_m)=\del_{mn},~~~~~~~~~~f(\xi_n)=a_n.}
We call  the states $\xi_n$ as \textbf{stationary states}, the numbers $a_n$ as \textbf{eigenvalues}  and the functions $u_n$ as \textbf{eigenfunctions} of the observable $f$. We denote this observable by $(f,a_n,u_n,\xi_n)$ and call it a \textbf{quantum function}.\\  
\textbf{Principle 3.} The dynamics of the system is given by the  Hamilton equation of a Hamiltonian $H$ which is a quantum function.

We name the values $(u_1(\xi),u_2(\xi),u_3(\xi),\cd)$ as the \textbf{quantum coordinates} of the state $\xi\in M$ \textbf{relative to the observable} $f$. Thus we realize that in spite of working on a finite-dimensional space of states $M$, we are assigning infinite-dimensional coordinates to each state in $M$. Thus we are recovering the infinite-dimensional aspect of standard quantum mechanics in our theory. In other words, we have an \textbf{absolute phase space} for states which is the finite-dimensional manifold $M$ and we also have a \textbf{relative phase space} for sates   relative to any given observable which may be infinite-dimensional.

\section{Recovering Standard Quantum Mechanics}
In this section we recover standard quantum theory through our theory. Namely, for each quantum function over phase space $M$ we  assign a complex Hilbert space and a Hermitian linear operator with the same eigenvalues and  we recover also the Schrodinger equation.

\begin{thm}
For any quantum function $f=\sum_na_n|u_n|^2$ with stationary states $\xi_n$ over phase space $M$ we associate a complex Hilbert space $\ch$ and a Hermitian linear operator $\tilde{f}$ such that $a_n$ are eigenvalues of $\tilde{f}$ and corresponding eigenvectors  $|\psi_n\teg$ form an orthonormal basis of $\ch$ and we assign a map
\eqn{}{\Phi:M\fl\ch,~~~~~~~~~~\Phi(\xi):=\sum_nu_n(\xi)|\psi_n\teg} which satisfies in 
\eqn{ihbarfPhi}{i\hbar\{f,\Phi\}=\tilde{f}\Phi,}
\eqn{f}{f=\textless\Phi|\tilde{f}|\Phi\textgreater,}
\eqn{Phixi}{|\Phi(\xi)|=1,~~~~~~~~~~\fa\xi\in M} and
\eqn{Phixin}{\Phi(\xi_n)=|\psi_n\teg.}Moreover, if $f=H$ is the Hamiltonian of the system then the solutions $\xi(t)$ of the Hamilton equation for this Hamiltonian are mapped  to the solutions $|\psi(t)\teg=\Phi(\xi(t))$ of the Schrodinger equation for the Hamiltonian $\tilde{H}$ operator.
\end{thm}
\B We relabel the stationary states $\xi_n$ by symbols $|\psi_n\teg$ and define $\ch$ to be the unique Hilbert space which has these vectors as  its orthonormal basis and also we define $\tilde{f}$ to be the unique Hermitian linear operator on $\ch$ with eigenvalues $a_n$ and its corresponding eigenvectors are $|\psi_n\teg$. Then equations (\ref{ihbarfPhi}), (\ref{f}), (\ref{Phixi}) and (\ref{Phixin}) are equivalent with the equations (\ref{ff}), (\ref{ihbarfun}), (\ref{sumnun2}) and (\ref{unxim}). For the last assertion we have $i\hbar\frac{d}{dt}|\psi\teg=i\hbar d_\xi\Phi(\dot{\xi})=i\hbar\{H,\Phi\}(\xi)=\tilde{H}\Phi(\xi)=\tilde{H}|\psi(t)\teg.$ \A 
\section{Construction of Quantum Functions from Quantum Operators}
\begin{thm}
Given a Hilbert space $\ch$, a Hermitian linear operator $A$ with eigenvalues $a_n$ and eigenvectors $|\psi_n\teg$ which form an orthonormal basis of $\ch$, a  map $\Phi:M\fl\ch$ over a phase space, some states $\xi_n\in M$ such that $|\Phi(\xi)|=1,\fa\xi\in M$, $\Phi(\xi_n)=|\psi_n\teg$ and 
\eqn{QFE}{i\hbar\{\textless\Phi|A|\Phi\textgreater,\Phi\}=A\Phi} then  the function
\eqn{}{\check{A}:=\textless\Phi|A|\Phi\textgreater}is a quantum function with stationary states $\xi_n$, eigenvalues $a_n$ whose corresponding  eigenvectors $u_n$ are the coefficients of the expansion of $\Phi$ \eqn{}{\Phi(\xi):=\sum_nu_n(\xi)|\psi_n\teg.}
\end{thm}
The proof is obvious. This  theorem suggests a method for construction of quantum functions; by solving the equation (\ref{QFE}).

\textbf{Conclusion}. We could translate principles of standard quantum mechanics from the language of  infinite-dimensional complex Hilbert spaces to the language of finite-dimensional real symplectic (Poisson) manifolds. There is one exception; Born rule which is the subject of our future research, inshallah.

\end{document}